\documentclass[11pt]{article}
\usepackage{fullpage}
\usepackage{amsmath}
\usepackage{amssymb}
\usepackage{amsfonts}
\usepackage{epsfig}
\usepackage[dvips]{color}

\newcounter{algoctr}

\newif\ifnotesw\noteswtrue
   {\ifnotesw\marginpar[\hfill\(\top\)]{\(\top\)}\fi}%
   {\ifnotesw\marginpar[\hfill\(\bot\)]{\(\bot\)}\fi}

\newcommand{\mnote}[1]%
    {\ifnotesw\marginpar%
        [{\scriptsize\begin{minipage}[t]{\marginparwidth}
        \raggedleft#1%
                        \end{minipage}}]%
        {\scriptsize\begin{minipage}[t]{\marginparwidth}
        \raggedright#1%
                        \end{minipage}}%
    \fi}

\newcommand{\ignore}[1]{}

\newcommand{\ie}{{\it i.e. }}

\newcommand{\etal}{{\it et al.~}}

\newsavebox{\given}
\savebox{\given}[1em]{\rule[-1.5ex]{.2mm}{4ex}}

\newtheorem{theorem}{Theorem}
\newtheorem{corollary}[theorem]{Corollary}

\newcommand{\blackslug}{\rule{7pt}{7pt}}

\newcommand{\iverson}[1]{\lbrack\!\lbrack #1 \rbrack\!\rbrack}

\newcommand{\qed}{\hfill{\setlength{\fboxsep}{0pt}
\framebox[7pt]{\rule{0pt}{7pt}}}}

\renewcommand{\notin}{\ifmmode \not\in \else $\not\in$ \fi}
\newlength{\thislabel}
\newcommand{\labsize}[1]{\settowidth{\thislabel}{#1}}

\newcommand{\prf}{\par\noindent{\sl Proof } \hspace{.01 in}}
\newcommand{\zo}{\{0,1\}}

\newcommand{\bra}[1]{\langle #1 |}
\newcommand{\ket}[1]{| #1 \rangle}
\newcommand{\braket}[2]{\langle #1 | #2 \rangle}
\newcommand{\ketbra}[2]{| #1 \rangle \langle #2 |}

\newcommand{\tstar}{t^{\star}}
\newcommand{\Int}{\mathbb{Z}}

\newcommand{\ajoin}[1]{+_{\mbox{\tiny #1}}}
\newcommand{\icgh}{\textsc{ICG}_{n}(\{1,n/2\})}
\newcommand{\icgq}{\textsc{ICG}_{n}(\{1,n/4\})}
\newcommand{\half}[1]{\lfloor #1/2 \rfloor}
\newcommand{\avg}[1]{\langle #1 \rangle}
\newcommand{\sqb}[1]{[ #1 ]}

\title{
Perfect state transfer, integral circulants and join of graphs
} 
\author{
{Ricardo Javier Angeles-Canul}\footnote{Centro de Investigaci\'{o}n en Matem\'{a}ticas, Universidad Aut\'{o}noma del Estado de Hidalgo, Pachuca, Hidalgo, Mexico. email: richywhitedragon@gmail.com}
\and
{Rachael M. Norton}\footnote{Dept. Mathematics, Bowdoin College, Brunswick, Maine, U.S.A. email: rnorton@bowdoin.edu}
\and
{Michael C. Opperman}\footnote{Dept. Mathematics and Computer Science, Clarkson University, Potsdam, New York, U.S.A. email: oppermmc@clarkson.edu}
\and
{Christopher C. Paribello}\footnote{Dept. Mathematics and Computer Science, Clarkson University, Potsdam, New York, U.S.A. email: paribecc@clarkson.edu}
\and
{Matthew C. Russell}\footnote{Dept. Mathematics, Taylor University, Upland, Indiana, U.S.A. email: matthew\_russell@taylor.edu}
\and
{Christino Tamon}\footnote{Dept. Mathematics and Computer Science, Clarkson University, Potsdam, New York, U.S.A. email: tino@clarkson.edu. Corresponding author.}
}
\ignore{
\author[1]{Ricardo Javier Angeles Canul}
\author[2]{Rachael Norton}
\author[3]{Michael Opperman}
\author[3]{Christopher Paribello}
\author[4]{Matthew Russell}
\author[3]{Christino Tamon}
\affil[1]{Universidad Aut\'{o}noma del Estado de Hidalgo, Mexico}
\affil[2]{Bowdoin College}
\affil[3]{Clarkson University}
\affil[4]{Taylor University}
}
\date{\today}

\begin{document}
\bibliographystyle{plain}
\maketitle

\begin{abstract}
We propose new families of graphs which exhibit quantum perfect state transfer.
Our constructions are based on the join operator on graphs, its circulant generalizations, 
and the Cartesian product of graphs. We build upon the results of Ba\v{s}i\'{c} \etal \cite{bps09,bp09}
and construct new integral circulants and regular graphs with perfect state transfer. 
More specifically, we show that the integral circulant $\textsc{ICG}_{n}(\{2,n/2^{b}\} \cup Q)$ 
has perfect state transfer, where $b \in \{1,2\}$, $n$ is a multiple of $16$ and 
$Q$ is a subset of the odd divisors of $n$.
Using the standard join of graphs, 
we also show a family of double-cone graphs which are non-periodic but exhibit perfect state transfer. 
This class of graphs is constructed by simply taking the join of the empty two-vertex graph with 
a specific class of regular graphs. This answers a question posed by Godsil \cite{godsil08}.

\vspace{.035in}
\end{abstract}



\section{Introduction}

In quantum information systems, the transfer of quantum states from one location to another is an important 
feature. The problem is to find an arrangement of $n$ interacting qubits in a network which allows perfect 
transfer of any quantum state over various distances.
The network is typically described by a graph where the vertices represent the location of the qubits
and the edges represent the pairwise coupling of the qubits. The graph has two special vertices labeled
$a$ and $b$ which represent the input (source) and output (target) qubits, respectively. 
In most cases of interest, it is required that perfect state transfer be achieved without dynamic control 
over the interactions between the qubits. These are the so-called permanently coupled (unmodulated) spin networks.

We may view the perfect state transfer problem in the context of quantum walks on graphs \cite{fg98,k06}.
Here the initial state of the quantum system is described by the unit vector on vertex $a$.
To achieve perfect transfer to vertex $b$ at time $t$, the quantum walk amplitude of the system at time $t$ 
on vertex $b$ must be of unit magnitude. That is, to obtain perfect transfer or unit fidelity, we require
\begin{equation}
|\bra{b}e^{-itA_{G}}\ket{a}| = 1,
\end{equation}
where $A_{G}$ is the adjacency matrix of the underlying graph $G$. The main goal here is to characterize graph
structures which allow such perfect state transfer.

Christandl \etal \cite{cdel04} showed that the Cartesian products of paths of length two or three possess
perfect state transfer between antipodal vertices -- vertices at maximum distance from each
other. They also noted that paths of length four or larger do not possess perfect state transfer if the 
edges are weighted equally. But, Christandl \etal \cite{cddekl05} showed that a layered path-like graph
of diameter $n$ has perfect state transfer, for any $n$.
More recently, Bernasconi \etal \cite{bgs08} gave a complete characterization 
for graphs from the hypercube family. They proved that perfect state transfer on the generalized $n$-cube 
is possible at time $t = \pi/2$ between ``antipodal'' vertices. Here, ``antipodal'' depends on the particular 
sequence that defines the generalized $n$-cube. 

Saxena \etal \cite{sss07}, Tsomokos \etal \cite{tpvh08}, and Ba\v{s}i\'{c} \etal \cite{bps09} studied perfect state 
transfer in integral circulant graphs. 
Tsomokos \etal \cite{tpvh08} showed perfect state transfer in the class of cross polytope or cocktail party graphs.
Ba\v{s}i\'{c} \etal \cite{bps09} completely characterized perfect state transfer on unitary Cayley graphs
which are equivalent to the integral circulants 
$\textsc{ICG}_{n}(\{1\})$ 
whose arc lengths must be relatively prime to $n$. 
They proved that $K_{2}$ and $C_{4}$ are the only unitary Cayley graphs with perfect state transfer.
Recently, Ba\v{s}i\'{c} and Petkovi\'{c} \cite{bp09} proved that the integral circulants 
$\icgq$ and $\icgh$, for $n$ divisible by $8$, have perfect state transfer. 
In the latter family of integral circulants, we have an example of graphs with perfect state transfer
between non-antipodal vertices -- vertices which are not at maximum distance from each other.
This answers a question posed by Godsil \cite{godsil08}.
In this paper, we construct new integral circulants with perfect state transfer 
by utilizing the join and Cartesian product of graphs.

\begin{figure}[t]
\begin{center}
\epsfig{file=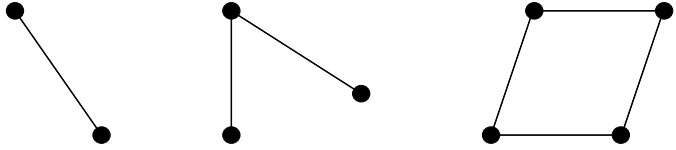, height=.75in, width=3.5in}
\caption{Small perfect state transfer graphs. 
From left to right: (a) $K_{2}$; (b) $P_{3}$; (c) $C_{4}$.}
\end{center}
\end{figure}

First, we generalize the graph join to an operation we call the {\em circulant join} $G \ajoin{C} G$ 
between a circulant graph $G$ and a Boolean circulant matrix $C$. This operation allows us to interpolate
between the standard join $G+G$ and the bunkbed (hypercube) operator $K_{2} \oplus G$ and, under certain
conditions, will produce new circulant graphs.
We recover the Cartesian product $K_{2} \oplus G$ by taking $C = I$, 
and the standard join $G+G$ (where all edges between vertices from the distinct copies of $G$ are present)
by taking $C = J$ (the all-one matrix).
If $G$ has perfect state transfer at time $\tstar$, then so does $G \ajoin{C} G$ at time $\tstar$
provided $\cos(\tstar\sqrt{C^{T}C})=\pm I$.
Moreover, $G \ajoin{C} G$ is a circulant graph whenever $C$ is a palindrome circulant; 
that is, the sequence which defines $C$ is a palindrome.
This allows us to construct new families of circulants with perfect state transfer,
namely, $\textsc{ICG}_{n}(\{2,n/2^{b}\} \cup Q)$, $b \in \{1,2\}$, where 
$n$ is divisible by $16$ and $Q$ is a subset of the odd divisors of $n$.
This expands the class of known integral circulants which exhibit perfect state transfer.

Next, we study graph operators that preserve perfect state transfer. A known example is the Cartesian
product of graphs as observed by Christandl \etal \cite{cdel04} in the $n$-fold Cartesian products of
paths of length two and three, namely, $K_{2}^{\oplus n}$ and $P_{3}^{\oplus n}$.
First, we note that this observation can be generalized to Cartesian products of different perfect state 
transfer graphs $\bigoplus_{k} G_{k}$ assuming these graphs have the same perfect state transfer times. 
We also prove closure properties of the $m$-fold self-join $\sum_{k=1}^{m} G$ of a graph $G$ with itself. 
In part, this is one generalization of the standard join $G+G$ we consider in this work. Using these results, 
we construct new graphs with perfect state transfer, such as 
$G^{\oplus m}$ and $G^{+m}$, where $G$ is one of $\textsc{ICG}_{n}(\{1,n/2^b\})$
or $\textsc{ICG}_{2n}(\{2,n/2^{b-1}\} \cup Q)$, where $n$ is a multiple of $8$, $b \in \{1,2\}$, 
and $Q$ is any subset of the odd divisors of $n$.
These new graphs, however, are not necessarily circulants. 

Finally, we consider the join of two arbitrary regular graphs.
Bose \etal \cite{bcms09} studied perfect state transfer on the complete graphs $K_{m}$ in the
so-called XYZ interaction model. Here, the quantum walk evolves according to the Laplacian of the
underlying graph instead of the adjacency matrix (the XY model). They show that, although $K_{m}$
does not have perfect state transfer, the double-cone $\overline{K}_{2} + K_{m-2}$ does. 
The latter graph is obtained from $K_{m}$ by removing an edge, say $(a,b)$, and
perfect state transfer occurs between $a$ and $b$.
We study a generalization of their construction by considering the join $G + H$ of two arbitrary regular graphs. 
We show that the existence of perfect state transfer on $G+H$ can be reduced to its existence in $G$
along with some additional conditions on the sizes and regularities of $G$ and $H$. These conditions
are independent of the internal structures of the graphs. 

Using this result, we construct a family of double-cone {\em non}-periodic graphs with perfect state transfer 
which answers a question posed by Godsil \cite{godsil08}. We also study the double-cone graphs 
$\overline{K}_{2} + G$, $K_{2} + G$, for any $n$-vertex $k$-regular graph $G$. 
We derive sufficient conditions on $n$ and $k$ which allow perfect state transfer 
between the two special vertices. This complements results found in \cite{bcms09} for the Laplacian model.
Our constructions involving $K_{2} + G$ also showed that perfect state transfer between non-antipodal vertices
is possible. As in the case of $\icgh$, this answers Godsil's other question \cite{godsil08}.

Our work heavily exploits the spectral properties of the underlying graphs and their matrices.
It is also based on the number-theoretic tools used to characterize integral circulants. 
A more complete treatment of this beautiful connection between circulants, number theory 
and graph theory can be found in earlier works by So \cite{s05}, Saxena \etal \cite{sss07}, 
and Ba\v{s}i\'{c} \etal \cite{bps09,bp09}.


\section{Preliminaries}

For a logical statement $\mathcal{S}$, the Iversonian $\iverson{\mathcal{S}}$ 
is $1$ if $\mathcal{S}$ is true and $0$ otherwise.
Let $\mathbb{Z}_{n}$ denote the additive group of integers $\{0,\ldots,n-1\}$ modulo $n$.
We use $I$ and $J$ to denote the identity and all-one matrices, respectively; we use
$X$ to denote the Pauli-$\sigma_{X}$ matrix.

The graphs $G=(V,E)$ we study are finite, simple, undirected, and connected. 
The adjacency matrix $A_{G}$ of a graph $G$ is defined as $A_{G}[u,v] = \iverson{(u,v) \in E}$.
A graph is called {\em integral} if its adjacency matrix has only integer eigenvalues.
A graph $G$ is {\em circulant} if its adjacency matrix $A_{G}$ is circulant.
A circulant matrix $A$ is completely specified by its first row, 
say $[a_{0}, a_{1}, \ldots, a_{n-1}]$, and is defined as 
$A[j,k] = a_{k-j \pmod{n}}$, where $j,k \in \Int_{n}$:
\begin{equation}
A = \begin{bmatrix}
    a_{0} & a_{1} & \ldots & a_{n-1} \\
    a_{n-1} & a_{0} & \ldots & a_{n-2} \\
    \vdots & \vdots & \ldots & \vdots \\
    a_{1} & a_{2} & \ldots & a_{0} 
	\end{bmatrix} 
\end{equation} 
Note that $a_{0} = 0$, since our graphs are simple, and $a_{j} = a_{n-j}$, since our graphs are undirected. 
The best known families of circulant graphs include the complete graphs $K_{n}$ and cycles $C_{n}$.

Alternatively, a circulant graph $G = (V,E)$ can be specified by a subset $S \subseteq \Int_{n}$, 
where $(j,k) \in E$ if $k-j \in S$. Thus, $S$ defines the set of {\em edge distances} between adjacent vertices.
In this case, we write $G = G(n,S)$. 
We will assume that $S$ is closed under taking inverses, namely, if $d \in S$, then $-d \in S$.
For a divisor $d$ of $n$, let $G_{n}(d) = \{k : \gcd(n,k)=d, 1 \le k < n\}$. 
It was proved by So \cite{s05} that a circulant $G(n,S)$ is integral if and only if 
$S = \bigcup_{d \in D} G_{n}(d)$, for some subset $D$ of $D_{n}$,
where $D_{n} = \{d : d | n, 1 \le d < n\}$ is the set of divisors of $n$. 
That is, a circulant is integral if its edge distances are elements of $G_{n}(d)$, $d \in D$, 
for some subset $D \subseteq D_{n}$. We denote this family of integral circulants as 
$\textsc{ICG}_{n}(D)$
(following the notation used in \cite{bps09}).

\begin{figure}[t]
\begin{center}
\epsfig{file=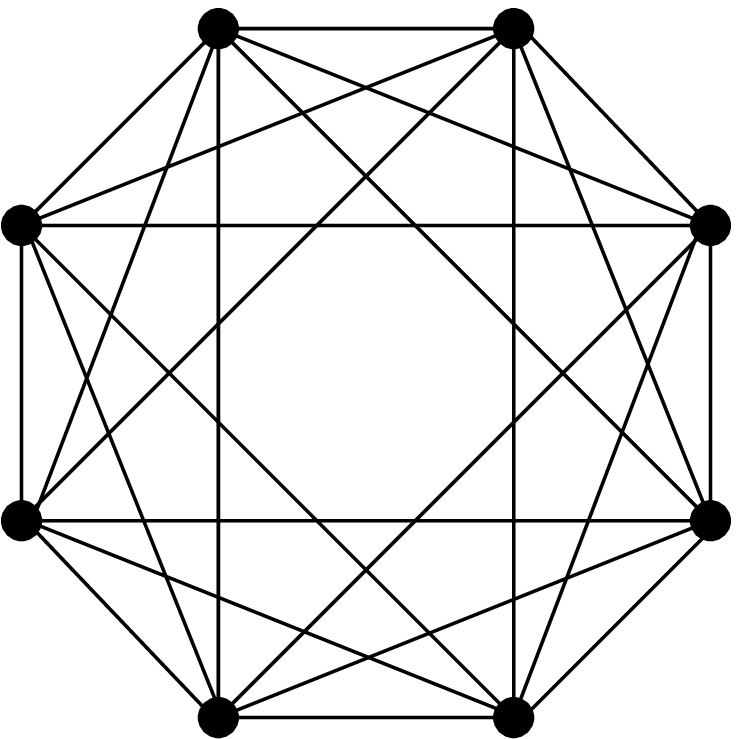, height=1.25in, width=1.25in}
\hspace{.5in}
\epsfig{file=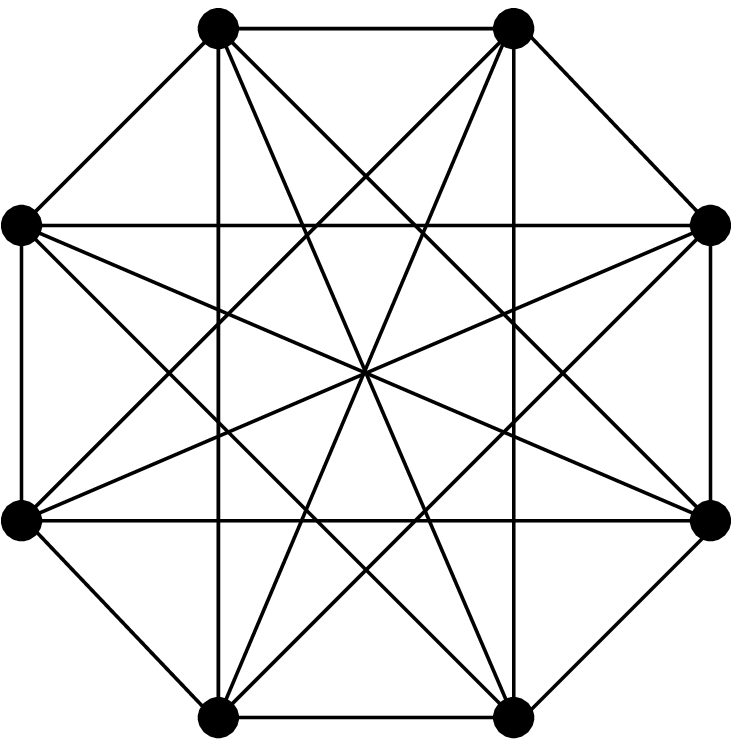, height=1.25in, width=1.25in}
\caption{Integral circulants with perfect state transfer. 
From left to right: (a) $\textsc{ICG}_{8}(\{1,2\})$; (b) $\textsc{ICG}_{8}(\{1,4\})$.
Perfect state transfer occurs from $x$ to $x+4$ at time $\pi/2$ in both graphs (see \cite{bp09}).
}
\end{center}
\end{figure}

All circulant graphs $G$ are diagonalizable by the Fourier matrix $F$ whose columns $\ket{F_{k}}$
are defined as $\braket{j}{F_{k}} = \omega_{n}^{jk}/\sqrt{n}$, where $\omega_{n} = \exp(2\pi i/n)$.
\ignore{
\begin{equation}
F = \frac{1}{\sqrt{n}}
    \begin{bmatrix} 
	1 & 1 & 1 & \ldots & 1 \\ 
	1 & \omega_{n} & \omega_{n}^{2} & \ldots & \omega_{n}^{n-1} \\ 
	1 & \omega_{n}^{2} & \omega_{n}^{4} & \ldots & \omega_{n}^{2(n-1)} \\ 
	\vdots & \vdots & \vdots & \ldots & \vdots \\ 
	1 & \omega_{n}^{n-1} & \omega_{n}^{2(n-1)} & \ldots & \omega_{n}^{(n-1)^{2}} 
	\end{bmatrix} 
\end{equation} 
}
In fact, we have $FAF^{\dagger} = \sqrt{n} \cdot diag(FA_{0})$, for any circulant $A$, 
where $A_{0}=A\ket{0}$ is the first column of $A$. 
This shows that the eigenvalues of $A$ are given by 
\begin{equation} \label{eqn:circulant-eigenvalue} 
\lambda_{j} = \sum_{k=0}^{n-1} a_{n-k} \ \omega_{n}^{jk}. 
\end{equation}

The Cartesian product $G \oplus H$ of graphs $G$ and $H$ is a graph whose adjacency matrix is
$I \otimes A_{H} + A_{G} \otimes I$. The {\em join} $G + H$ of graphs $G$ and $H$ is defined
as $\overline{G+H} = \overline{G} \cup \overline{H}$; that is, we connect all vertices of $G$
with all vertices of $H$. The adjacency matrix of $G+H$ is given by 
$\begin{bmatrix} A_{G} & J \\ J & A_{H} \end{bmatrix}$, with the appropriate dimensions on
the two all-one $J$ matrices.
For more background on algebraic graph theory, we refer the reader to the monographs 
by Biggs and by Godsil and Royle \cite{biggs,gr} as well as to the survey article
by Schwenk and Wilson \cite{sw78}.

For a graph $G=(V,E)$, let $\ket{\psi(t)} \in \mathbb{C}^{|V|}$ be a time-dependent amplitude vector 
over $V$. Then the continuous-time quantum walk on $G$ is defined using Schr\"{o}dinger's equation as
\begin{equation}
\ket{\psi(t)} = e^{-it A_{G}} \ket{\psi(0)},
\end{equation}
where $\ket{\psi(0)}$ is the initial amplitude vector (see \cite{fg98}). 
Further background on quantum walks on graphs can be found in the survey by Kendon \cite{k06}.
We say $G$ has {\em perfect state transfer} from vertex $a$ to vertex $b$ at time $\tstar$ if
\begin{equation} \label{eqn:pst}
|\bra{b}e^{-i\tstar A_{G}}\ket{a}| = 1,
\end{equation}
where $\ket{a}$, $\ket{b}$ denote the unit vectors corresponding to the vertices $a$ and $b$,
respectively. The graph $G$ has perfect state transfer if there exist vertices $a$ and $b$ in $G$
and a time $\tstar$ so that (\ref{eqn:pst}) is true. 
Also, we call a graph $G$ {\em periodic} if for any state $\ket{\psi}$, there is a time $\tstar$
so that $|\bra{\psi}e^{-it A_{G}}\ket{\psi}| = 1$.


\section{Circulant Joins}

In this section, we describe a new graph operator which preserves perfect state transfer.
For a $n$-vertex graph $G$ and a $n \times n$ Boolean matrix $C$, 
define the circulant join $\mathcal{G} = G \ajoin{C} G$ as a graph
whose adjacency matrix is 
\begin{equation}
A_{\mathcal{G}} = \begin{bmatrix} A_{G} & C \\ C^{T} & A_{G} \end{bmatrix}.
\end{equation}
That is, we take two copies of $G$ and connect vertices from the corresponding copies using the matrix $C$. 
Here, we do not require that $C$ be the adjacency matrix of a graph.
This generalizes the join $G + G = G \ajoin{J} G$ and the bunkbed $K_{2} \oplus G = G \ajoin{I} G$.
For these self-join constructions of a graph $G$ with itself, where there are two copies of $G$, 
if $u$ is a vertex of $G$, then we denote $(u,s)$, $s \in \zo$, as the vertex $u$ in the $s$-th copy of $G$.

\begin{theorem} \label{thm:circulant-join}
Let $C$ be a $n \times n$ circulant matrix.
If $G$ is a $n$-vertex circulant graph with perfect state transfer from $a$ to $b$ at time $\tstar$, 
then the circulant join $G \ajoin{C} G$ has perfect state transfer from 
vertex $(a,0)$ to vertex $(b,s)$, $s \in \zo$, at time $\tstar$ provided that
\begin{equation}
\left[\cos(\tstar\sqrt{B})\right]^{1-s} \left[\sin(\tstar\sqrt{B})B^{-1/2}C^{T}\right]^{s} = \pm I
\end{equation}
where $B = C^{T}C$, and $B^{-1}$ exists whenever $s=1$.
Moreover, $G \ajoin{C} G$ is a circulant graph if $C$ is a palindrome circulant matrix, 
where $c_{j} = c_{n-1-j}$, for $j=0,\ldots,n-1$.
\end{theorem}
\prf
Note that the adjacency matrix of $\mathcal{G} = G \ajoin{C} G$ can be rearranged as 
\begin{equation}
\mathcal{C}_{A} = \left(C \otimes \ketbra{0}{1} + C^{T} \otimes \ketbra{1}{0}\right) + A_{G} \otimes I_{2}.
\end{equation}
It is clear that $C$ and $C^{T}$ commute since they are both circulants.
Next, observe that
\begin{equation}
\left[C \otimes \ketbra{0}{1} + C^{T} \otimes \ketbra{1}{0}\right]^{\ell}
=
\left\{\begin{array}{ll}
	B^{k} \otimes I_{2} & \mbox{ if $\ell=2k$ } \\
	B^{k}C \otimes \ketbra{0}{1} + B^{k}C^{T} \otimes \ketbra{1}{0} & \mbox{ if $\ell=2k+1$ }
	\end{array} \right.
\end{equation}
In the above equation, notice that the even or odd powers vanish depending on $s$:
\begin{equation}
\bra{b}\bra{0}
\left[B^{k}C \otimes \ketbra{0}{1} + B^{k}C^{T} \otimes \ketbra{1}{0}\right]
\ket{a}\ket{0}
	= 0.
\end{equation}
and
\begin{equation}
\bra{b}\bra{1}
B^{k} \otimes I_{2}
\ket{a}\ket{0}
	= 0.
\end{equation}
\ignore{
Using this, we see that
\begin{equation}
\bra{b}\bra{s} e^{-it\mathcal{C}_{A}} \ket{a}\ket{0}
	= \left\{\begin{array}{ll}
	\bra{b} \cos(\tstar\sqrt{B}) e^{-it A_{G}} \ket{a} & \mbox{ if $s=0$} \\ 
	-i \bra{b} \sin(\tstar\sqrt{B}) B^{-1/2}C^{T} e^{-it A_{G}} \ket{a} & \mbox{ if $s=1$}
	\end{array}\right.
\end{equation}
}
Thus, perfect state transfer in $\mathcal{G}$ can be reduced to perfect state transfer in $G$ as follows:
\begin{eqnarray}
\bra{b,s}e^{-i\tstar \mathcal{C}_{A}}\ket{a,0} 
	& = & \bra{b}\bra{s}e^{-i\tstar(C \otimes \ketbra{0}{1} + C^{T} \otimes \ketbra{1}{0})}
			e^{-i\tstar(A_{G} \otimes I_{2})}\ket{a}\ket{0} \\
	& = & \bra{b}\bra{s}e^{-i\tstar(C \otimes \ketbra{0}{1} + C^{T} \otimes \ketbra{1}{0})}
			(e^{-i\tstar A_{G}} \otimes I_{2})\ket{a}\ket{0} \\
\label{eqn:cos-sin}
	& = &
	\left\{\begin{array}{ll}
	\bra{b} \cos(\tstar\sqrt{B}) e^{-i\tstar A_{G}} \ket{a} & \mbox{ if $s=0$} \\ 
	-i \bra{b} \sin(\tstar\sqrt{B}) B^{-1/2}C^{T} e^{-i\tstar A_{G}} \ket{a} & \mbox{ if $s=1$}
	\end{array}\right.
\end{eqnarray}
This proves the first claim.

To see that $\mathcal{G}$ is a circulant graph if $C$ is a palindrome circulant matrix, 
we view the adjacency matrix $\mathcal{C}_{A}$ as an ``interweaving'' of $A_{G}$ with $C$ and $C^{T}$ as follows:
\begin{equation} \label{eqn:interlace-join}
\mathcal{C}_{A} = 
	\begin{bmatrix}
	a_{0}   & \avg{c_{0}}   & a_{1}   & \avg{c_{1}}  & a_{2}   & \avg{c_{2}}  & \ldots & a_{n-1} & \avg{c_{n-1}} \\
	\sqb{c_{0}}   & a_{0}   & \sqb{c_{n-1}} & a_{1}  & \sqb{c_{n-2}} & a_{2}  & \ldots & \sqb{c_{1}}   & a_{n-1} \\
	a_{n-1} & \avg{c_{n-1}} & a_{0}   & \avg{c_{0}}  & a_{1}   & \avg{c_{1}}  & \ldots & a_{n-2} & \avg{c_{n-2}} \\
	\sqb{c_{1}}   & a_{n-1} & \sqb{c_{0}}   & a_{0}  & \sqb{c_{n-1}} & a_{1}  & \ldots & \sqb{c_{2}}   & a_{n-2} \\
	\vdots  & \vdots  & \vdots  & \vdots & \vdots  & \vdots & \ldots & \vdots  & \vdots  \\
	a_{1}   & \avg{c_{1}}   & a_{2}   & \avg{c_{2}}  & a_{3}   & \avg{c_{3}}  & \ldots & a_{0}   & \avg{c_{0}}   \\
	\sqb{c_{n-1}} & a_{1}   & \sqb{c_{n-2}} & a_{2}  & \sqb{c_{n-3}} & a_{3}  & \ldots & \sqb{c_{0}}   & a_{0}
	\end{bmatrix}
\end{equation}
where $A_{G} = (a_{j})$, $C = (\avg{c_{j}})$, and $C^{T} = (\sqb{c_{j}})$.
We have distinguished the elements of $C$ and $C^{T}$ by using $\avg{c}$ and $\sqb{c}$, respectively. 
Now, applying $c_{j}=c_{n-1-j}$, for $j=0,\ldots,n-1$, 
\ignore{
we have
\begin{equation} \label{eqn:interlace-join}
\mathcal{C}_{A} = 
	\begin{bmatrix}
	a_{0}   & c_{0}   & a_{1}   & c_{1}  & a_{2}  & c_{2}  & \ldots & a_{n-1} & c_{n-1} \\
	c_{n-1} & a_{0}   & c_{0}   & a_{1}  & c_{1}  & a_{2}  & \ldots & c_{n-2} & a_{n-1} \\
	a_{n-1} & c_{n-1} & a_{0}   & c_{0}  & a_{1}  & c_{1}  & \ldots & a_{n-2} & c_{n-2} \\
	c_{n-2} & a_{n-1} & c_{n-1} & a_{0}  & c_{0}  & a_{1}  & \ldots & c_{n-3} & a_{n-2} \\
	\vdots  & \vdots  & \vdots  & \vdots & \vdots & \vdots & \ldots & \vdots  & \vdots  \\
	a_{1}   & c_{1}   & a_{2}   & c_{2}  & a_{3}  & c_{3}  & \ldots & a_{0}   & c_{0}   \\
	c_{n-1} & a_{1}   & c_{1}   & a_{2}  & c_{2}  & a_{3}  & \ldots & c_{n-1} & a_{0}
	\end{bmatrix}
\end{equation}
where 
}
it is clear that the above is a circulant matrix; hence, $\mathcal{G}$ is a circulant graph.
\ignore{ 
The interlacing in (\ref{eqn:interlace-join}) can also be viewed equivalently as
\begin{equation}
\mathcal{C}_{A} = 
	\begin{bmatrix}
	B_{0}     & B_{1}     & B_{2}  & \ldots & B_{2}^{T} & B_{1}^{T} \\
	B_{1}^{T} & B_{0}     & B_{1}  & \ldots & B_{3}^{T} & B_{2}^{T} \\ 
	B_{2}^{T} & B_{1}^{T} & B_{0}  & \ldots & B_{4}^{T} & B_{3}^{T} \\
	\vdots    & \vdots    & \vdots & \ldots & \vdots    & \vdots    \\
	B_{2}     & B_{3}     & B_{4}  & \ldots & B_{0}     & B_{1}     \\
	B_{1}     & B_{2}     & B_{3}  & \ldots & B_{1}^{T} & B_{0}
	\end{bmatrix}
	\ \ \ \mbox{ where }
	B_{j} = \begin{bmatrix} a_{j} & c_{j} \\ c_{j-1} & a_{j} \end{bmatrix}.
\end{equation}
which shows a block circulant structure. This completes the proof.
}
\qed\\

As corollaries to the above theorem, we show the conditions for which the hypercube Cartesian product and
the join of a perfect state transfer graph with itself preserves the perfect state transfer property.

\begin{corollary} \label{cor:cartesian}
If $G$ is an $n$-vertex circulant that has perfect state transfer from $a$ to $b$ at time 
$\tstar \in (2\mathbb{Z}+1)\frac{\pi}{2}$, then the bunkbed $K_{2} \oplus G$ has perfect state
transfer from $(a,0)$ to $(b,1)$, where $(a,0)$ denotes vertex $a$ in the first copy
of $G$ and $(b,1)$ denotes vertex $b$ in the second copy of $G$.
\end{corollary}
\prf
Since $K_{2} \oplus G = G \ajoin{I} G$, the eigenvalues of the connection matrix are $\mu_{k}=1$ for all $k$. 
By Theorem \ref{thm:circulant-join}, we require $\sin(\tstar I_{n})=\pm I_{n}$ for perfect state transfer.
This is equivalent to $\sin(\tstar)=\pm 1$ which is satisfied whenever $\tstar \in (2\mathbb{Z}+1)\frac{\pi}{2}$.
\qed\\

\begin{corollary} \label{cor:join}
If $G$ is an $n$-vertex circulant graph that has perfect state transfer from $a$ to $b$ at time 
$\tstar$, then so does the circulant graph $G + G$ provided $n\tstar \in 2\pi\mathbb{Z}$.
\end{corollary}
\prf
Since $G + G = G \ajoin{J} G$, by Theorem \ref{thm:circulant-join} we require 
$\cos(\tstar\Lambda)=I_{n}$, where $\Lambda = diag(n,0,\ldots,0)$. This is equivalent to
requiring $\cos(n\tstar)=1$ which is satisfied when $n\tstar \in 2\pi\mathbb{Z}$.
\qed\\


\par\noindent{\em Remark}: It can be shown that for $n = 2^{u}$, where $u \ge 3$, the only
Boolean circulant matrices $C$ that yield a circulant graph $G \ajoin{C} G$, for an $n$-vertex $G$,
are the trivial matrices, namely, $C \in \{I_{n},J_{n},O_{n}\}$, where $O_{n}$ is the $n \times n$
all-zero matrix. In the next theorem, we show that for $n$ that is a multiple of $8$, if $n$ has a
non-trivial odd divisor, then there exist integral circulant graphs 
$\textsc{ICG}_{2n}(D)$, for $|D| \ge 3$, with perfect state transfer which are obtained 
from non-trivial circulant joins.\\ 

\par\noindent{\em Notation}: For an integer $m$, let $\overline{D}_{m} = \{d : d|m, 1 \le d \le m\}$ 
be the set of all divisors of $m$. Also, for an integer $k$ and a set $A \subseteq \mathbb{Z}$, we use 
$kA$ to denote $\{ka: a\in A\}$.

\begin{theorem} \label{thm:general-nontrivial-circulant-join}
Let $n = 2^{u}m$, where $u \ge 3$ and $m \ge 3$ is an odd number.
Suppose that $G = \textsc{ICG}_{n}(D)$, for $D = \{1,n/4\}$ or $D = \{1,n/2\}$.
For any subset $Q \subset \overline{D}_{m}$, there is a Boolean circulant matrix 
$C \not\in \{I_{n},J_{n},O_{n}\}$ so that 
\begin{equation}
G \ajoin{C} G = \textsc{ICG}_{2n}(2D \cup Q)
\end{equation}
has perfect state transfer from $0$ to $n/2$ in $G$ at time $\tstar = \pi/2$,
\end{theorem}
\prf
For $q \in Q$, let $N(q) = \{r \in \overline{D}_{m}: r/q \textnormal{ is an odd prime }\}$.
Define, for $j=0,\ldots,n-1$,
\begin{equation}
c_{j}(q) = \iverson{2j+1\equiv 0\hspace{-.125in}\pmod{q} 
	\wedge \forall r \in N(q): 2j+1\not\equiv 0\hspace{-.125in}\pmod{r}}.
\end{equation}
Now let $c_{j}(Q) = \sum_{q \in Q} c_{j}(q)$. Note that $c_{j}(q)$'s are disjoint, since at most
one index $j$ will satisfy $c_{j}(q)=1$ for $q \in Q$.
The Boolean circulant $C$ is defined by the following first row
\begin{equation}
C = [c_{0}(Q) \ \ldots \ c_{n-1}(Q)].
\end{equation}
To see that $C$ is a palindrome, note that $2(n-1-j)+1 \equiv 0\pmod{q}$ is equivalent to 
$2j+1 \equiv 0\pmod{q}$, for any $q \in \overline{D}_{m}$.

Since the integral circulants $\textsc{ICG}_{n}(\{1,n/2^{b}\})$, $b \in \{1,2\}$, 
have $\tstar = \pi/2$ as the perfect state transfer time, we will show that 
$\cos(|\lambda_{k}|\pi/2) = 1$, for all eigenvalues $\lambda_{k}$ of $C$, $k=0,\ldots,n-1$.
The eigenvalues of $C$ are given by 
\begin{eqnarray}
\lambda_{k} 
	& = & \sum_{j=0}^{n-1} c_{j}(Q) \omega_{n}^{-jk} 
	= \sum_{j=0}^{n-1} \sum_{q \in Q} c_{j}(q) \omega_{n}^{-jk} 
	= \sum_{q \in Q} \sum_{j=0}^{n-1} c_{j}(q) \omega_{n}^{-jk} \\
	& = & \sum_{q \in Q} 
		\left\{\sum_{\ell=0}^{\frac{n}{q}-1} \omega_{n}^{-(\half{q}+\ell q)k} 
			- \sum_{r \in N(q)} 
				\sum_{\ell=0}^{\frac{n}{r}-1} \omega_{n}^{-(\half{r}+\ell r)k} \right\} \\
	& = & \sum_{q \in Q} 
		\left\{\frac{n}{q} \omega_{n}^{-\half{q}k} \iverson{k \equiv 0\hspace{-.125in}\pmod{n/q}}
			- \sum_{r \in N(q)} 
				\frac{n}{r} \omega_{n}^{-\half{r}k} 
				\iverson{k \equiv 0\hspace{-.125in}\pmod{n/r}} \right\}
\end{eqnarray}
Now, we consider the exponent in the term $\omega_{n}^{-\half{q}k}$, where $k$ satisfies 
$k \equiv 0\pmod{n/q}$. Thus, there is an integer $\kappa$ so that $qk = \kappa n$ or 
$(2\half{q}+1)k = \kappa n$. Note that $\kappa$ depends on $k$.
After rearranging, we get $-\half{q}k = \frac{k}{2}-\kappa\frac{n}{2}$. Thus,
\begin{equation}
\omega_{n}^{-\half{q}k} 
	= \omega_{n}^{k/2}(\omega_{n}^{-n/2})^{\kappa}
	= \omega_{n}^{k/2}(-1)^{\kappa}.
\end{equation}
Thus, we have
\begin{equation}
\lambda_{k} \in \omega_{n}^{k/2}(-1)^{\kappa} 2^{u}\mathbb{Z}.
\end{equation}
It is clear now that $\cos(|\lambda_{k}|\pi/2) = \cos(2\pi\mathbb{Z}) = 1$, for all $k = 0,\ldots,n-1$.
So, by Theorem \ref{thm:circulant-join}, $G \ajoin{C} G$ has perfect state transfer at time $\pi/2$.

Now, we show that if $G = \textsc{ICG}_{n}(D)$, then $G \ajoin{C} G = \textsc{ICG}_{2n}(2D \cup Q)$,
where $2D = \{2d: d \in D\}$.
Let $B$ be the circulant adjacency matrix of $ G \ajoin{C} G$ defined by the sequence $[b_{0},\ldots,b_{2n-1}]$.
From the ``interweaving'' property of $B$ in (\ref{eqn:interlace-join}), we know that for
$k \in \{0,\ldots,2n-1\}$
\begin{equation}
b_{k} = \left\{\begin{array}{ll}
		a_{k/2} & \mbox{ if $k$ is even } \\
		c_{\lfloor k/2 \rfloor} & \mbox{ if $k$ is odd }
		\end{array} \right.
\end{equation}
We consider two cases based on whether $k$ is even or odd.
For $k$ odd, we have
\begin{eqnarray}
b_{k} & = & c_{\half{k}} \\
	& = & \iverson{\exists q \in Q: 2\half{k}+1\equiv 0\hspace{-.125in}\pmod{q} 
			\wedge \forall r \in N(q): 2\half{k}+1\not\equiv 0\hspace{-.125in}\pmod{r}} \\
	& = & \iverson{\exists q\ \in Q: k\equiv 0\hspace{-.125in}\pmod{q} 
			\wedge \forall r \in N(q): k\not\equiv 0\hspace{-.125in}\pmod{r}} \\
	& = & \iverson{\exists q \in Q: \gcd(k,2n) = q},
\end{eqnarray}
whereas for $k$ even, we have
\begin{equation}
b_{k} = a_{k/2} = \iverson{\gcd(k/2,n) \in D} = \iverson{\gcd(k,2n) \in 2D}.
\end{equation}
This proves the claim by taking $D = \{1,n/4\}$ or $D = \{1,n/2\}$.
\qed\\

\begin{corollary} \label{cor:icg-join}
For $n=2^{u}$, for $u \ge 3$, the integral circulant graphs 
$\textsc{ICG}_{2n}(\{1,2,n/2\})$ and $\textsc{ICG}_{2n}(\{1,2,n\})$ have perfect state transfer from $0$ to $n$ 
at time $\tstar = \pi/2$.
\end{corollary}
\prf
Take the self-joins of $\icgq$ and $\icgh$ which have perfect state transfer as shown by 
Ba\v{s}i\'{c} and Petkovi\'{c} \cite{bp09}.
\qed\\

\par\noindent In the next corollary, we consider the class of circulant permutation matrices $C$.
These matrices are defined by specifying a unit vector (with a one in a single entry and zeros elsewhere)
as their first row, and they satisfy $C^{T}C = I$. We show that if $G$ has perfect state transfer from $a$
to $b$ at time $\tstar$, then so does $G \ajoin{C} G$ from $a$ in the first copy of $G$ to $Cb$
(image of vertex $b$ under the permutation $C$) in the second copy of $G$ at time $\tstar$.

\begin{corollary} \label{cor:permutation-circulant}
Let $G$ be a $n$-vertex graph and let $C$ be a $n \times n$ circulant permutation matrix. 
If $G$ has perfect state transfer from vertex $a$ to $b$ at time $\tstar$, then
$G \ajoin{C} G$ has perfect state transfer from vertex $(a,0)$ to $(Cb,1)$ at time $\tstar$,
where $Cb$ is the image of vertex $b$ under the permutation $C$.
\end{corollary}
\prf
Let $\mathcal{G} = G \ajoin{C} G$.
Since $C$ is a permutation matrix, we have $C^{T}C = I$. 
Applying (\ref{eqn:cos-sin}) in the proof of Theorem \ref{thm:circulant-join}, we get
\begin{equation}
\bra{b}\bra{1}e^{-itA_{\mathcal{G}}}\ket{a}\ket{0} = -i\bra{b}C^{T}e^{-itA_{G}}\ket{a}.
\end{equation}
which can be rearranged as
$\bra{Cb}\bra{1}e^{-itA_{\mathcal{G}}}\ket{a}\ket{0} = -i\bra{b}e^{-itA_{G}}\ket{a}$.
\qed\\

\par\noindent{\em Remark}: The above corollary shows that if there is perfect state transfer in $G$
from vertex $a$ to vertex $b$ then there is perfect state transfer from $(a,0)$ to any vertex $(c,1)$
in $G \ajoin{C} G$. This is achieved by choosing a permutation matrix $C$ so that $Cb = c$.

\section{Cartesian Products and Self-Joins}

We show that a heterogeneous Cartesian product of perfect state transfer graphs has
perfect state transfer. This generalizes the observations on the hypercube $K_{2}^{\oplus n}$ 
and the Cartesian product $P_{3}^{\oplus n}$ of paths of length three (see \cite{cdel04}).

\begin{figure}[t]
\begin{center}
\epsfig{file=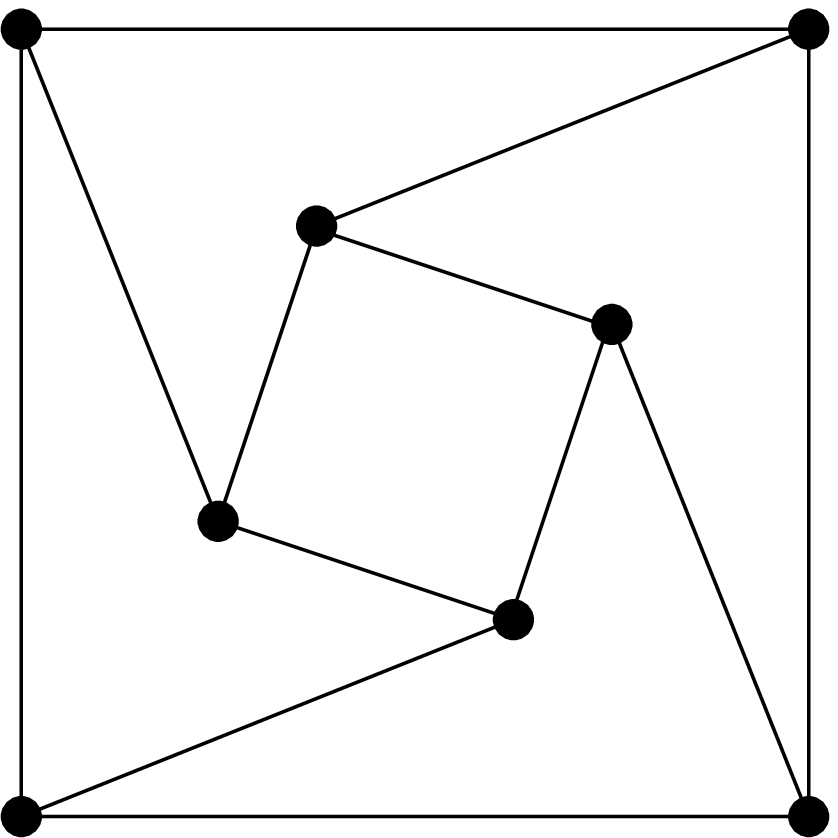, height=1.25in, width=1.25in}
\hspace{.5in}
\epsfig{file=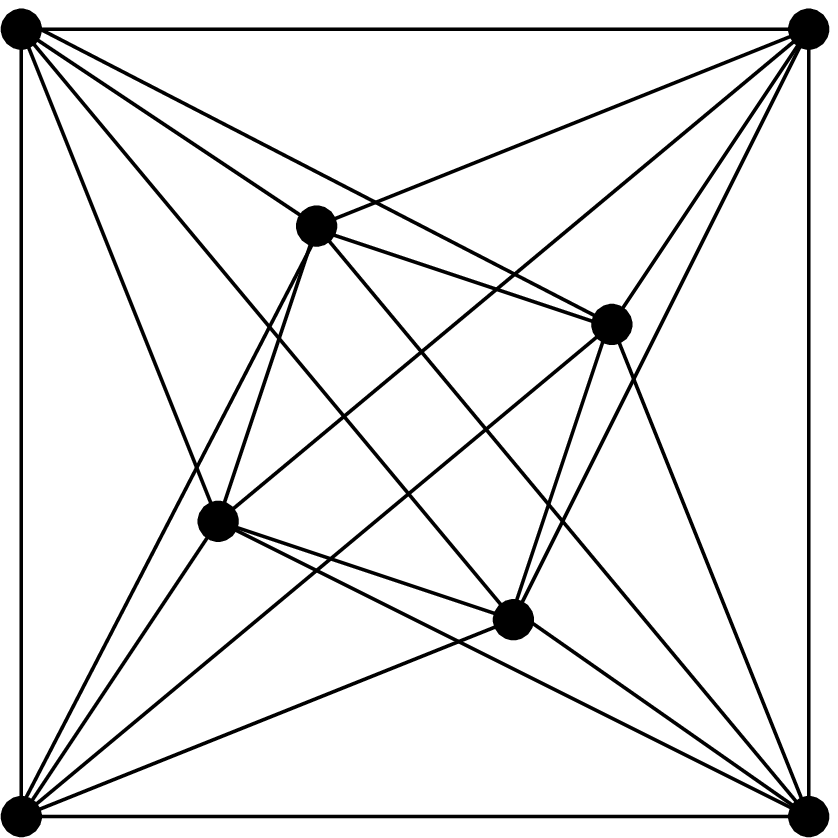, height=1.25in, width=1.25in}
\caption{Standard circulant joins on $C_{4}$.
From left to right: (a) Cartesian bunkbed $C_{4} \ajoin{I} C_{4}$; (b) Self-join $C_{4} \ajoin{J} C_{4}$}
\end{center}
\end{figure}

\begin{theorem} \label{thm:cartesian}
For $j=1,\ldots,m$, the graph $G_{j}$ has perfect state transfer from $a_{j}$ to $b_{j}$ at time $\tstar$ 
if and only if 
$\mathcal{G} = \bigoplus_{j=1}^{m} G_{j}$ has perfect state transfer 
from $(a_{1},\ldots,a_{m})$ to $(b_{1},\ldots,b_{m})$ at time $\tstar$.
\end{theorem}
\prf
We prove the claim for $m=2$. Then,
\begin{eqnarray}
\bra{b_{1},b_{2}}e^{-itA_{G_{1} \oplus G_{2}}}\ket{a_{1},a_{2}}
	& = & \bra{b_{1},b_{2}}e^{-it(I \otimes A_{G_{2}} +  A_{G_{1}} \otimes I)}\ket{a_{1},a_{2}} \\
	& = & \bra{b_{1}}\bra{b_{2}}e^{-it(I \otimes A_{G_{2}})}e^{-it(A_{G_{1}} \otimes I)}\ket{a_{1}}\ket{a_{2}} \\
	& = & \bra{b_{1}}\bra{b_{2}}(I \otimes e^{-it A_{G_{2}}})(e^{-it A_{G_{1}}} \otimes I)\ket{a_{1}}\ket{a_{2}} \\
	& = & \bra{b_{1}}e^{-it A_{G_{1}}}\ket{a_{1}}\bra{b_{2}}e^{-it A_{G_{2}}}\ket{a_{2}}.
\end{eqnarray}
This shows that $G_{1} \oplus G_{2}$ has perfect state transfer from $(a_{1},a_{2})$ to $(b_{1},b_{2})$
at time $\tstar$ if and only if 
$G_{1}$ has perfect state transfer from $a_{1}$ to $b_{1}$ at time $\tstar$
{\em and}
$G_{2}$ has perfect state transfer from $a_{2}$ to $b_{2}$ at time $\tstar$.
The general claim follows easily by induction.
\qed\\

\begin{corollary}
For any $m$ and $n$ so that $n \equiv 0\pmod{8}$, the family of graphs $\bigoplus_{k=1}^{m} G_{k}$, 
where $G_{k} \in \{\textsc{ICG}_{n}(\{1,n/2^b\}), \textsc{ICG}_{2n}(\{2,n/2^{b-1}\} \cup Q)\}$,
$b \in \{1,2\}$ and $Q$ is a subset of the set of odd divisors of $n$,
has perfect state transfer from vertex $0$ to $n/2$ at time $\tstar = \pi/2$.
\end{corollary}
\prf
Follows from Theorem \ref{thm:cartesian}, the results of Ba\v{s}i\'{c} \etal \cite{bp09},
and Theorem \ref{thm:circulant-join}.
\qed\\

\par\noindent We also show that the $m$-fold join of a perfect state transfer graph preserves perfect state
transfer under certain conditions. Denote $G^{+m}$ as the $m$-fold self-join $\sum_{j=1}^{m} G$. 

\begin{theorem} \label{thm:self-join}
Let $G$ be an $n$-vertex regular graph. For $m \ge 1$, the existence of perfect state transfer in $G^{+m}$ 
between vertices $a$ and $b$ (in the same copy of $G$) can be reduced to its existence in $G$ as follows:
\begin{equation} \label{eqn:self-join}
\bra{0,b}e^{-it A_{G^{+m}}}\ket{0,a} = \bra{b}e^{-it A_{G}}\ket{a} + 
	\left[\frac{(m-1)(e^{itn}-1) + e^{-it(m-1)n}-1}{mn}\right] \bra{1_{n}}e^{-it A_{G}}\ket{a},
\end{equation}
where $A_{G}$ is the adjacency matrix of $G$ and $\ket{1_{n}}$ is the all-one column vector of length $n$.
\end{theorem}
\prf
The adjacency matrix of $G^{+m}$ is given by $I_{m} \otimes A_{G} + K_{m} \otimes J_{n}$, where $K_{m}$ is
the adjacency matrix of the complete graph on $m$ vertices. 
First note that $J_{n}^{\ell} = n^{\ell-1}J_{n}$, if $\ell \ge 1$, and $J_{n}^{0} = I_{n}$.
Also, notice that $A_{G}$ commutes with $J_{n}$ since $G$ is a regular graph.
Moreover, $K_{m} = J_{m} - I_{m}$. Thus, using the binomial theorem, we get
\begin{equation}
K_{m}^{\ell} = \frac{1}{m}\left\{(-1)^{\ell}(mI_{m} - J_{m}) + (m-1)^{\ell} J_{m}\right\}.
\end{equation}
\ignore{
\begin{eqnarray}
K_{m}^{\ell} & = & (J_{m} - I_{m})^{\ell} \\
	& = & \sum_{j=0}^{\ell} {\ell \choose j} J_{m}^{j} (-I_{m})^{\ell-j} \\
	& = & (-1)^{\ell}I_{m} + \sum_{j=1}^{\ell} {\ell \choose j} m^{j-1} J_{m} (-1)^{\ell-j} \\
	& = & (-1)^{\ell}I_{m} + \frac{J_{m}}{m} \sum_{j=1}^{\ell} {\ell \choose j} m^{j} (-1)^{\ell-j} \\
	& = & \frac{(-1)^{\ell}}{m}(mI_{m} - J_{m}) + \frac{J_{m}}{m} (m-1)^{\ell} \\
	& = & \frac{1}{m}\left\{(-1)^{\ell}(mI_{m} - J_{m}) + (m-1)^{\ell} J_{m}\right\}.
\end{eqnarray}
}
Therefore,
\begin{eqnarray} 
e^{-it(K_{m} \otimes J_{n})} 
	& = & \sum_{\ell=0}^{\infty} \frac{(-it)^{\ell}}{\ell !} K_{m}^{\ell} \otimes J_{n}^{\ell} \\
	& = & I_{m} \otimes I_{n} + \sum_{\ell=1}^{\infty} \frac{(-it)^{\ell}}{\ell !} K_{m}^{\ell} \otimes J_{n}^{\ell} \\
	& = & I_{m} \otimes I_{n} + \frac{1}{n} \sum_{\ell=1}^{\infty} \frac{(-it)^{\ell}}{\ell !} K_{m}^{\ell} \otimes (n^{\ell}J_{n}) \\
	\label{eqn:Km-join}
	& = & I_{m} \otimes I_{n} + 
		\frac{(e^{itn}-1)}{mn} (mI_{m}-J_{m}) \otimes J_{n} +
		\frac{(e^{-it(m-1)n}-1)}{mn} J_{m} \otimes J_{n}
\end{eqnarray}
We can now analyze the quantum walk amplitude from vertex $a$ to $b$ (in the same copy of $G$).
We get
\begin{eqnarray}
\bra{0,b}e^{-it A_{G^{+m}}}\ket{0,a} 
	& = & \bra{0}\bra{b} e^{-it(K_{m} \otimes J_{n})} e^{-it(I_{m} \otimes A_{G})} \ket{0}\ket{a} \\
	& = & \bra{0}\bra{b} e^{-it(K_{m} \otimes J_{n})} (I_{m} \otimes e^{-it A_{G}}) \ket{0}\ket{a} \\
	& = & \bra{0}\bra{b} e^{-it(K_{m} \otimes J_{n})} (\ket{0} \otimes e^{-it A_{G}}\ket{a})
\end{eqnarray}
Expanding the second term using (\ref{eqn:Km-join}) and multiplying the two terms on the left, we get
\begin{equation}
\left(\bra{0}\bra{b} + \frac{(e^{itn}-1)}{mn}(m\bra{0}-\bra{1_{m}}) \otimes \bra{1_{n}} + 
		\frac{(e^{-it(m-1)n}-1)}{mn} \bra{1_{m}} \otimes \bra{1_{n}}\right)
		(\ket{0} \otimes e^{-it A_{G}}\ket{a}).
\end{equation}
Finally, combining this with the last term, we arrive at
\begin{equation}
\bra{b}e^{-it A_{G}}\ket{a} + 
	\frac{1}{mn}\left[(e^{itn}-1)(m-1) + e^{-it(m-1)n} - 1\right] \bra{1_{n}}e^{-it A_{G}}\ket{a},
\end{equation}
which proves the claim.
\qed\\

\begin{corollary}
For any $m \ge 1$ and $n \equiv 0\pmod{8}$,
the family of graphs $G^{+m}$, 
where $G \in \{\textsc{ICG}_{n}(\{1,n/2^{b}\}), \textsc{ICG}_{2n}(\{2,n/2^{b-1}\} \cup Q)\}$, 
with $n \equiv 0\pmod{8}$, $b \in \{1,2\}$, and a subset $Q$ of the odd divisors of $n$,
has perfect state transfer between vertices $0$ and $n/2$ (in the same copy of $G$).
\end{corollary}
\prf
By Theorem \ref{thm:self-join}, to achieve perfect state transfer in $G^{+m}$, it suffices to 
have perfect state transfer in $G$ at time $\tstar$ and have 
$e^{i\tstar n} = 1$. 
By the results of Ba\v{s}i\'{c} \etal \cite{bp09} and by Theorem \ref{thm:general-nontrivial-circulant-join},
the integral circulant graphs stated in the claim
have perfect state transfer from vertex $0$ to vertex $n/2$ at time $\tstar = \pi/2$. 
Therefore, it suffices to have $n \in 4\mathbb{Z}$. 
Since $n \equiv 0\pmod{8}$, this holds for any $m$.
\qed\\

\begin{corollary}
For any $m \ge 1$ and $n \ge 2$, the family of graphs $Q_{n}^{+m}$, 
where $Q_{n}$ is the binary $n$-dimensional hypercube,
has perfect state transfer between antipodal vertices in the same copy of $Q_{n}$.
\end{corollary}
\prf
Bernasconi \etal \cite{bgs08} proved that $Q_{n}$ has perfect state transfer between its antipodal vertices at time
$\tstar = \pi/2$. Note that the number of vertices of $Q_{n}$ is $N=2^{n}$.
Using Theorem \ref{thm:self-join}, it suffices to set $e^{i\tstar N}=1$ or $N \equiv 0\pmod{4}$. 
This is always true since $N=2^{n}$, with $n \ge 2$.
\qed


\section{Join of Regular Graphs}

We show that the existence of perfect state transfer in a join of two arbitrary regular graphs can be reduced to 
perfect state transfer in one of the graphs along with certain additional constraints on the sizes and degrees
of the graphs. These conditions are independent of the internal structures of the graphs.

\begin{theorem} \label{thm:binary-join}
Let $G$ be an $m$-vertex $k_{G}$-regular graph and let $H$ be an $n$-vertex $k_{H}$-regular graph.
Suppose that $a$ and $b$ are two vertices in $G$.
Then, 
\begin{equation} \label{eqn:pst-join-reduction}
\bra{b}e^{-it A_{G+H}}\ket{a}
	= \bra{b} e^{-it A_{G}} \ket{a} 
	+
	\frac{e^{-it k_{G}}}{m}
	\left\{e^{it\delta/2} \left[ \cos\left(\frac{\Delta t}{2}\right) 
	- i\left(\frac{\delta}{\Delta}\right)\sin\left(\frac{\Delta t}{2}\right) \right] - 1 \right\}
\end{equation}
where $\delta = k_{G}-k_{H}$ and $\Delta = \sqrt{\delta^{2} + 4mn}$.
\end{theorem}
\prf
Let $a,b$ be two vertices of $G$. 
Then,
\begin{equation}
\bra{b}e^{-it A_{G}}\ket{a} = 
	\bra{b}\left\{\sum_{k=0}^{m-1} \ket{u_{k}}\bra{u_{k}} e^{-it\lambda_{k}}\right\} \ket{a} 
\end{equation}
where $\lambda_{k}$ and $\ket{u_{k}}$ are the eigenvalues and eigenvectors of $A_{G}$,
for $k=0,\ldots,m-1$. We assume $\ket{u_{0}}$ is the all-one eigenvector (that is orthogonal 
to the other eigenvectors) with eigenvalue $\lambda_{0} = k_{G}$. By the same token, let
$\kappa_{\ell}$ and $\ket{v_{\ell}}$ be the eigenvalues and eigenvectors of $A_{H}$, for
$\ell = 0,\ldots,n-1$. Also, $\ket{v_{0}}$ is the all-one eigenvector (with eigenvalue $\kappa_{0} = k_{H}$) 
which is orthogonal to the other eigenvectors $\ket{v_{\ell}}$, $\ell \neq 0$.

Let $\mathcal{G} = G + H$. Note that the adjacency matrix of $\mathcal{G}$ is
\begin{equation}
A_{\mathcal{G}} = \begin{bmatrix} A_{G} & J_{m \times n} \\ J_{n \times m} & A_{H} \end{bmatrix}.
\end{equation}
Let $\delta = k_{G}-k_{H}$.
The eigenvalues and eigenvectors of $A_{\mathcal{G}}$ are given by the three sets:
\begin{itemize}
\item 
	For $k=1,\ldots,m-1$, let $\ket{u_{k},0_{n}}$ be a column vector formed by concatenating the 
	column vector $\ket{u_{k}}$ with the zero vector of length $n$. 
	Then, $\ket{u_{k},0_{n}}$ is an eigenvector with eigenvalue $\lambda_{k}$.

\item
	For $\ell=1,\ldots,n-1$, let $\ket{0_{m},v_{\ell}}$ be a column vector formed by concatenating 
	the zero vector of length $m$ with the column vector $\ket{v_{\ell}}$.
	Then, $\ket{0_{m},v_{\ell}}$ is an eigenvector with eigenvalue $\kappa_{\ell}$.

\item
	Let $\ket{\pm} = \frac{1}{\sqrt{L_{\pm}}}\ket{\alpha_{\pm},1_{n}}$ be a column vector formed
	by concatenating the vector $\alpha_{\pm}\ket{1_{m}}$ with the vector $\ket{1_{n}}$, where
	$\ket{1_{m}}$, $\ket{1_{n}}$ denote the all-one vectors of length $m$, $n$, respectively.
	Then, $\ket{\pm}$ is an eigenvector with eigenvalue $\lambda_{\pm} = k_{H} + m\alpha_{\pm}$.
	Here,
\begin{equation}
\alpha_{\pm} = \frac{1}{2m}(\delta \pm \Delta), \ \ \ 
L_{\pm} = m(\alpha_{\pm})^{2} + n.
\end{equation}

\end{itemize}
In what follows, we will abuse notation by using $\ket{a}$, $\ket{b}$ for both $G$ and $G+H$;
their dimensions differ in both cases, although it will be clear from context which version is used.
The quantum wave amplitude from $a$ to $b$ is given by
\begin{eqnarray} 
\bra{b}e^{-it A_{\mathcal{G}}}\ket{a}
	& = & \bra{b} e^{-it A_{\mathcal{G}}} \left\{\sum_{k=1}^{m-1}\braket{u_{k},0_{n}}{a} \ket{u_{k},0_{n}} + 
			\sum_{\pm} \frac{\alpha_{\pm}}{\sqrt{L_{\pm}}} \ket{\pm} \right\} \\
	& = & \bra{b} \left\{\sum_{k=1}^{m-1}\braket{u_{k}}{a} e^{-it\lambda_{k}}\ket{u_{k},0_{n}} + 
			\sum_{\pm} \frac{\alpha_{\pm}}{\sqrt{L_{\pm}}} e^{-it\lambda_{\pm}} \ket{\pm} \right\} \\
	& = & \sum_{k=1}^{m-1}\braket{b}{u_{k}}\braket{u_{k}}{a} e^{-it\lambda_{k}} + 
			\sum_{\pm} \frac{\alpha_{\pm}^{2}}{L_{\pm}} e^{-it\lambda_{\pm}} \\
\label{eqn:pst-join}
	& = & \bra{b}\left\{\sum_{k=0}^{m-1}\ket{u_{k}}\bra{u_{k}} e^{-it\lambda_{k}}\right\} \ket{a} 
			-\frac{e^{-it k_{G}}}{m} + 
			\sum_{\pm} \frac{\alpha_{\pm}^{2}}{L_{\pm}} e^{-it\lambda_{\pm}} \\
\label{eqn:pst-join2}
	& = & \bra{b} e^{-it A_{G}} \ket{a} 
			+ \sum_{\pm} \frac{\alpha_{\pm}^{2}}{L_{\pm}} e^{-it\lambda_{\pm}} 
			-\frac{e^{-it k_{G}}}{m}.
\end{eqnarray}
We analyze the second term next. Note that we have the following identities:
\begin{eqnarray}
\alpha_{+}\alpha_{-} & = & -(n/m) \\
\alpha_{+} + \alpha_{-} & = & \delta/m \\
L_{+}L_{-} & = & (n/m)\Delta^{2} \\
L_{+} + L_{-} & = & \Delta^{2}/m \\
(\alpha_{\pm})^{2}L_{\mp} & = & (n/m)L_{\pm} \\
\label{eqn:lambda-pm}
\lambda_{\pm} & = & (\hat{\delta} \pm \Delta)/2
\end{eqnarray}
where $\hat{\delta} = k_{G} + k_{H}$. 
Therefore, the summand in (\ref{eqn:pst-join2}) is given by
\begin{eqnarray}
\sum_{\pm} \frac{\alpha_{\pm}^{2}}{L_{\pm}} e^{-it\lambda_{\pm}} 
	& = & \frac{1}{m} e^{-it\hat{\delta}/2} \left[ \cos\left(\frac{\Delta t}{2}\right) 
				- i\left(\frac{\delta}{\Delta}\right)\sin\left(\frac{\Delta t}{2}\right) \right].
\end{eqnarray}
This yields
\begin{equation}
\bra{b}e^{-it A_{\mathcal{G}}}\ket{a}
	= \bra{b} e^{-it A_{G}} \ket{a} 
	+
	\frac{e^{-it k_{G}}}{m}
	\left\{e^{it\delta/2} \left[ \cos\left(\frac{\Delta t}{2}\right) 
	- i\left(\frac{\delta}{\Delta}\right)\sin\left(\frac{\Delta t}{2}\right) \right] - 1 \right\}
\end{equation}
which proves the claim.
\ignore{ 
So, 
we have the condition:
\begin{eqnarray}
\frac{e^{-it k_{G}}}{m}
	\left\{e^{it^{\star}\delta/2} \left[ \cos\left(\frac{t^{\star}\Delta}{2}\right) 
	- i\left(\frac{\delta}{\Delta}\right)\sin\left(\frac{t^{\star}\Delta}{2}\right) \right] - 1 \right\}
	= 0.
\end{eqnarray}
Therefore, for perfect state transfer to occur from $a$ to $b$ at time $\tstar$, it suffices to guarantee
\begin{equation}
\cos\left(\frac{\delta t^{\star}}{2}\right)\cos\left(\frac{\Delta t^{\star}}{2}\right) = 1.
\end{equation}
}
\qed\\

\begin{figure}[t]
\begin{center}
\epsfig{file=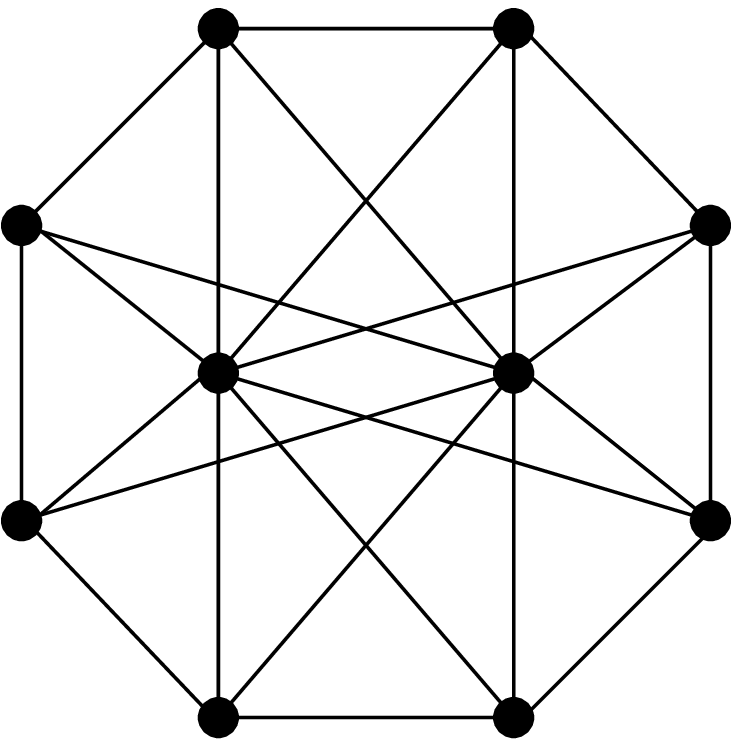, height=1.25in, width=1.25in}
\hspace{.5in}
\epsfig{file=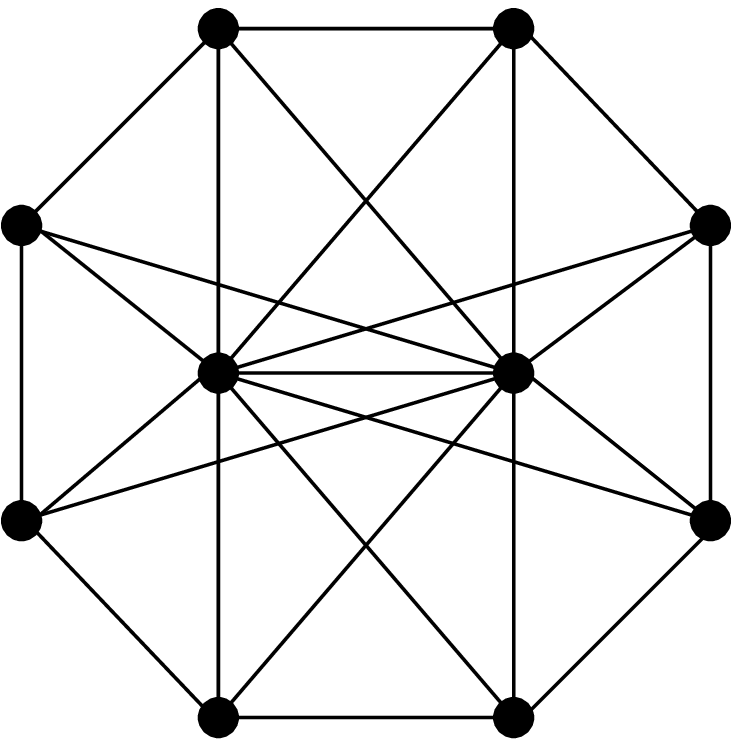, height=1.25in, width=1.25in}
\hspace{.5in}
\epsfig{file=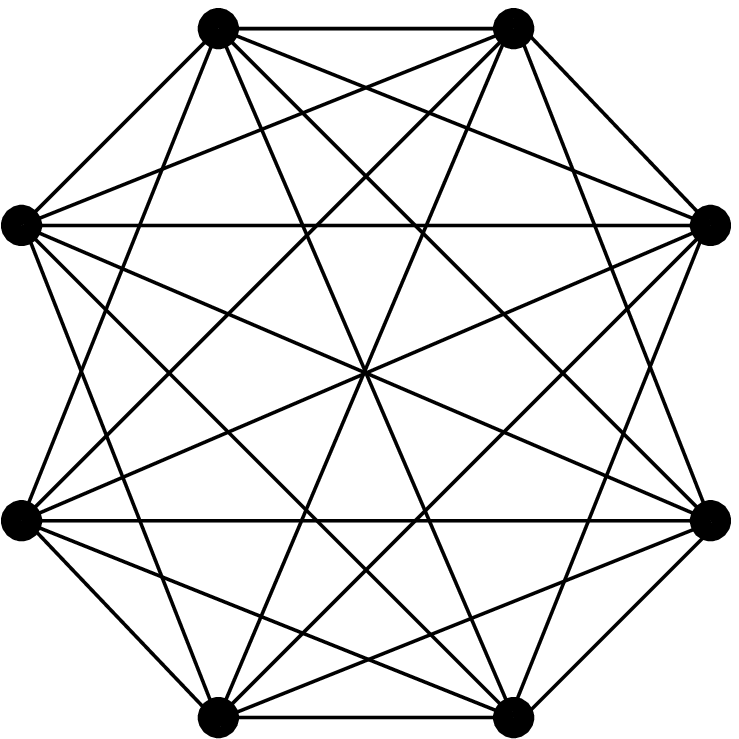, height=1.25in, width=1.25in}
\caption{Double Cones. From left to right: (a) $\overline{K}_{2} + C_{8}$; 
(b) $K_{2} + C_{8}$; (c) Cocktail Party (hyperoctahedral).
}
\label{fig:double-cone}
\end{center}
\end{figure}

\par\noindent In what follows, we describe several applications of Theorem \ref{thm:binary-join} to the
double-cones $\overline{K}_{2} + G$ and $K_{2} + G$ and also to the construction of a family of non-periodic
graphs with perfect state transfer. The existence of the latter family of graphs was one of the main
questions posed by Godsil \cite{godsil08}.
For a prime $p$, we denote $S_{p}(n)$ to be the largest non-negative integer $j$ so that $p^{j} | n$.

\begin{corollary} \label{cor:disconnected-double-cone}
For any $k$-regular graph $G$ on $n$ vertices, $\overline{K}_{2} + G$ has perfect state transfer between
the two non-adjacent vertices of $\overline{K}_{2}$ if
$\Delta = \sqrt{k^{2}+8n}$ is an integer and $k,\Delta \equiv 0\pmod{4}$
with $S_{2}(k) \neq S_{2}(\Delta)$.
\end{corollary}
\prf
Let $\mathcal{G} = \overline{K}_{2} + G$. 
By Theorem \ref{thm:binary-join}, since there is no transfer between the two vertices of $\overline{K}_{2}$,
we have
\begin{equation}
\bra{b}e^{-it A_{\mathcal{G}}}\ket{a} =
	\frac{1}{2}
	\left\{e^{-it k/2} \left[ \cos\left(\frac{\Delta t}{2}\right) 
		+ i\left(\frac{k}{\Delta}\right)\sin\left(\frac{\Delta t}{2}\right) \right] - 1 \right\}.
\end{equation}
To achieve unit magnitude, it is necessary and sufficient to require $\cos(kt/2)\cos(\Delta t/2) = -1$.
Note that setting $k=0$ (\ie $G$ is the empty graph on $n$ vertices) will result in this condition being
satisfied for $t=2 \pi / \Delta$.  This is, in a sense, a generalization of $P_3$, which has perfect state 
transfer~\cite{cdel04}.

Otherwise, assume $k = 2^{k_{0}}k_{1}$, where $k_{1}$ is odd; 
and $\Delta = 2^{d_{0}}d_{1}$, where $d_{1}$ is odd.
Since there is no transfer in $\overline{K}_{2}$, we may choose $t \in \mathbb{Q}\pi$ and 
require that $\Delta$ be an integer.
It is clear that $kt/2$ and $\Delta t/2$ must have opposite parities as multiples of $\pi$.
This implies $k_{0} \neq d_{0}$. If $k_{0} > d_{0}$, we have
\begin{equation}
n = \frac{1}{8}(\Delta^{2} - k^{2})
	= \frac{1}{8}(4^{d_{0}}d_{1}^{2}-4^{k_{0}}k_{1}^{2})
	= \frac{4^{d_{0}}}{8}(d_{1}^{2} - 4^{k_{0}-d_{0}}k_{1}^{2}).
\end{equation}
Since $(d_{1}^{2} - 4^{k_{0}-d_{0}}k_{1}^{2})$ is odd and $n$ is an integer, $8$ divides $4^{d_{0}}$ which
implies $d_{0} \ge 2$ and $k_{0} > 2$. A similar argument when $k_{0} < d_{0}$ shows that
$d_{0} > k_{0} \ge 2$. Thus, both $k$ and $\Delta$ are multiples of $4$.
\qed\\

\par\noindent{\em Remark}: 
Using $k \equiv 0 \pmod 4, n = k+2, \Delta = k+4$ satisfy the conditions of
Corollary~\ref{cor:disconnected-double-cone}, and thus the graph $\mathcal{G} = \overline{K}_{2} + G$
has perfect state transfer.  In this case, $\mathcal G$ can be represented by a type of circulant graph 
called a \emph{hyperoctahedral}, or \emph{cocktail-party}, graph~\cite{biggs}
(see Figure \ref{fig:double-cone}). 
These graphs are formed by removing $n/2+1$ disjoint edges from $K_{n+2}$.
This class of graphs, which is also called the class of cross polytope graphs,
was also studied by Tsomokos \etal \cite{tpvh08}.
\\

\par\noindent Next, we answer a question of Godsil \cite{godsil08} (Section 10, question (b))
by constructing an infinite family of non-periodic graphs with the perfect state transfer property.

\begin{corollary} \label{cor:cone-grid}
For $\ell \ge 2$, the family of double-cone graphs
$\overline{K}_{2} + (C_{2(2\ell-1)} \oplus C_{2\ell+1})$
is non-periodic and has perfect state transfer.
\end{corollary}
\prf
Let $G = C_{2(2\ell-1)} \oplus C_{2\ell+1}$, for $\ell \ge 2$.
Note that $G$ is a $k$-regular graph with $k=4$ and $n = 2(4\ell^{2}-1)$ vertices. 
Using the notation of Theorem \ref{thm:binary-join}, 
we have $\Delta = \sqrt{k^{2} + 8n} = 8\ell$.
The eigenvalues of $G$ are given by the sum of the eigenvalues of the two cycles:
\begin{equation}
\lambda(G) = \lambda(C_{2(2\ell-1)}) + \lambda(C_{2\ell+1}).
\end{equation}
Recall that the eigenvalues of an $n$-cycle are given by $2\cos(2\pi k/n)$, for $k=0,\ldots,n-1$. 
So, each cycle has $2$ (its degree) as its largest eigenvalue.
Thus, the sums of the cycle eigenvalues contain both integers and irrational numbers. 
For $n=5$ and $n \ge 7$, at least some of these values are irrational.
This is because the only rational values of $\cos((a/b)\pi)$, for $a,b \in \mathbb{Z}$,
are $\{0,\pm 1/2,\pm 1\}$ (see Corollary 3.12 in Niven \cite{niven}).
Note that $2(2\ell-1) \ge 5$ and $2\ell+1 \ge 5$ hold for $\ell \ge 2$, 
and that both expressions cannot equal $6$.

The eigenvalues of $\mathcal{G}=\overline{K}_{2} + G$ will then be all of the eigenvalues of
$G$ (except for 4), 0, and 
$\lambda_{\pm}=\frac{1}{2}(4 \pm 8\ell) = 2 \pm 4\ell$ 
by (\ref{eqn:lambda-pm}).
This means that $\mathcal{G}$ has a mixture of integral and irrational eigenvalues. 
By Lemma 4.1 in \cite{godsil08}, the graph $\mathcal{G}$ is non-periodic.
By Corollary \ref{cor:disconnected-double-cone}, 
since $\Delta = 8\ell$ is an integer and $S_{2}(k) \neq S_{2}(\Delta)$,
we know that $\mathcal{G}$ has perfect state transfer. 
This proves the claim.
\qed\\

\par\noindent{\em Remark}: 
Taking $\ell = 2$ in Corollary \ref{cor:cone-grid}, we get $\mathcal{G} = \overline{K}_{2} + C_{5} \oplus C_{6}$.
Again by Lemma 4.1 in \cite{godsil08}, $\mathcal{G}$ is non-periodic since its eigenvalues contain both integers
and irrational numbers. By Corollary \ref{cor:cone-grid}, we know it has perfect state transfer although it 
violates the {\em eigenvalue ratio} condition 
$(\lambda_{k}-\lambda_{\ell})/(\lambda_{r}-\lambda_{s}) \in \mathbb{Q}$, for $\lambda_{r} \neq \lambda_{s}$.
This is in contrast to Theorem 2.1 in \cite{godsil08} and to the mirror-symmetric networks in Section III 
from \cite{cddekl05}. Our double-cone construction is mirror-symmetric with respect to the
two vertices of $\overline{K}_{2}$.

\begin{corollary} \label{cor:connected-double-cone}
For any $n$-vertex $k$-regular graph $G$, let $\tilde{k} = k - 1$.
Then, $K_{2} + G$ has perfect state transfer between the two adjacent vertices of $K_{2}$ if
$\Delta = \sqrt{\tilde{k}^{2}+8n}$ is an integer and $\tilde{k},\Delta \equiv 0\pmod{8}$.
\end{corollary}
\prf
Let $\mathcal{G} = K_{2} + G$. 
By Theorem \ref{thm:binary-join}, since there is perfect state transfer between the two vertices of 
$K_{2}$ at time $\tstar = (2\mathbb{Z}+1)\frac{\pi}{2}$, we have
\begin{equation}
\bra{b}e^{-i\tstar A_{\mathcal{G}}}\ket{a} = \bra{b}e^{-i\tstar A_{G}}\ket{a}
	+
	\frac{e^{-i\tstar}}{2}
	\left\{e^{-i\tstar\tilde{k}/2} \left[ \cos\left(\frac{\tstar\Delta}{2}\right) 
		+ i\left(\frac{\tilde{k}}{\Delta}\right)\sin\left(\frac{\tstar\Delta}{2}\right) \right] - 1 \right\}.
\end{equation}
To achieve perfect state transfer in $\mathcal{G}$, it suffices to require
\begin{equation}
	\frac{e^{-i\tstar}}{2}
	\left\{e^{-i\tstar\tilde{k}/2} \left[ \cos\left(\frac{\tstar\Delta}{2}\right) 
		+ i\left(\frac{\tilde{k}}{\Delta}\right)\sin\left(\frac{\tstar\Delta}{2}\right) \right] - 1 \right\}
	= 0
\end{equation}
or equivalently, $\cos(\tstar\tilde{k}/2)\cos(\tstar\Delta/2) = 1$ with $\Delta \in \mathbb{Z}$.
Thus, it is sufficient to choose $\tilde{k}, \Delta \equiv 0\pmod{8}$ 
given $\tstar \in (2\mathbb{Z}+1)\frac{\pi}{2}$.
\qed\\

\par\noindent{\em Remark}: 
The above corollaries complement the results of Bose \etal \cite{bcms09} on $K_{2} + K_{m-2}$ 
and $\overline{K}_{2} + K_{m-2}$ in the XYZ (Laplacian) interaction model. They showed that 
$K_{m}$ has no perfect state transfer, but if we delete the edge $(a,b)$ then there is perfect 
state transfer between $a$ and $b$.


\section{Conclusion}

In this work, we studied perfect state transfer on quantum networks represented by graphs in the XY 
(adjacency) interaction model. 
Prior to our work, the only unweighted graphs known to have perfect state transfer were the cube-like 
networks \cite{bgs08}, the Cartesian product $P_{3}^{\oplus n}$ of paths of length three \cite{cdel04}, 
the path-like layered graph of diameter $n$ \cite{cddekl05}, and the integral circulant graphs 
$\textsc{ICG}_{n}(\{1,n/2^{b}\})$, for $n \equiv 0\pmod{8}$ and $b \in \{1,2\}$ \cite{bp09}. 
We described constructions of new families of graphs with perfect state transfer using graph operators which 
preserve this property. More specifically, we used the graph-theoretic join and its circulant generalizations
as well as the Cartesian product. Most of our results involved a {\em reduction} argument from the 
larger graph structure to the individual graphs with respect to the perfect state transfer property.

We generalized both the ``hypercube'' Cartesian product and the join graph operators by defining a so-called
{\em circulant join} $G \ajoin{C} G$ of two copies of a circulant graph $G$ and connecting them using 
a circulant matrix $C$. 
This allowed us to interpolate between the above two interesting constructions and produced a graph operator 
that preserves the circulant property. From this construction, we derived new families of circulant graphs 
with perfect state transfer, namely, $\textsc{ICG}_{n}(\{2,n/2^{b}\} \cup Q)$, $b \in \{1,2\}$, where $Q$
is a subset of the odd divisors of $n$.
This expanded the class of circulant graphs known to have perfect state transfer (see \cite{bps09,bp09}).

Then, we showed that the Cartesian product of different perfect state transfer graphs has perfect state 
transfer provided all of these graphs share the same transfer time. This generalized previous results for 
paths of length two and three \cite{cdel04}. For the $n$-fold self-join, we showed that the existence of 
perfect state transfer on $G^{+n}$ can be reduced to its existence in $G$ along with other conditions. 
These observations allowed us to construct new families of graphs with perfect state transfer, 
for example $\icgh^{+m}$ or $\bigoplus_{k=1}^{m} G_{k}$, where $G_{k}$ is any
of the known integral circulant and hypercubic graphs with perfect state transfer, for any integer $m$.

Finally, we considered the join $G + H$ of two arbitrary regular graphs. Again, we reduced the existence of 
perfect state transfer on $G+H$ to its existence in $G$ along with some conditions on the sizes and degrees of 
the two graphs. From this reduction, we constructed an interesting double-cone family of graphs 
$\overline{K}_{2}+G$ which are non-periodic but with perfect state transfer. 
This answered one of the main questions posed in Godsil \cite{godsil08}. 
This also complemented results in Bose \etal \cite{bcms09} on perfect state transfer in the complete graphs 
$K_{m}$ with a missing edge. Their results were stated in the XYZ (Laplacian) model while our results hold 
in the XY (adjacency) model.

It seems plausible that there is a characterization of perfect state transfer in integral circulants
$G(n,\bigcup_{d \in D} G_{n}(d))$, for any $D$, using the join operator. This will complement the 
characterization of unitary Cayley graphs where $D = \{1\}$ \cite{bps09}.
It would also be interesting to consider perfect state transfer in weighted graphs (see \cite{cdel04}), 
especially on unweighted graphs which are known to lack the property.
Finally, we find it curious that most of the graphs known with perfect state transfer achieve this
at time $\tstar = (2\mathbb{Z}+1)\pi/2$. This is true for the cube-like graphs \cite{bgs08} and for the integral 
circulants \cite{bp09}. The lone exceptions are paths of length three \cite{cdel04} and
the double-cones $\overline{K}_{2} + G$, for suitable chosen regular graphs $G$.


\section*{Acknowledgments}

We thank Garry Bowlin for helpful comments on the circulant join construction.
This research was supported in part by the National Science Foundation grant DMS-0646847
and also by the National Security Agency grant H98230-09-1-0098.



\begin{thebibliography}{000}

\bibitem{biggs}
Norman Biggs,
{\em Algebraic Graph Theory},
2nd edition, Cambridge University Press, 1993.

\bibitem{bcms09}
Sougato Bose, Andrea Casaccino, Stefano Mancini, and Simone Severini,
``Communication in XYZ All-to-All Quantum Networks with a Missing Link,''
{\em International Journal on Quantum Information} {\bf 7}(4):713-723, 2009.

\bibitem{bgs08}
Anna Bernasconi, Chris Godsil, and Simone Severini,
``Quantum Networks on Cubelike Graphs,''
{\em Physical Review A} {\bf 78}, 052320, 2008. 

\bibitem{bp09}
Milan Ba\v{s}i\'{c} and Marko Petkovi\'{c},
``Some classes of integral circulant graphs either allowing or not allowing perfect state transfer,''
{\em Applied Mathematics Letters}, in press.

\bibitem{bps09}
Milan Ba\v{s}i\'{c}, Marko Petkovi\'{c}, and Dragan Stefanovi\'{c},
``Perfect state transfer in integral circulant graphs,''
{\em Applied Mathematics Letters} {\bf 22}(7):1117-1121, 2009.

\bibitem{cdel04}
Matthias Christandl, Nilanjana Datta, Artur Ekert, Andrew Landahl,
``Perfect state transfer in quantum spin networks,''
{\em Physical Review Letters} {\bf 92}, 187902, 2004. 

\bibitem{cddekl05}
Matthias Christandl, Nilanjana Datta, Tony Dorlas, Artur Ekert, Alastair Kay, Andrew Landahl,
``Perfect transfer of arbitrary states in quantum spin networks,''
{\em Physical Review A} {\bf 71}, 032312, 2005. 

\bibitem{fg98}
Edward Farhi and Sam Gutmann,
``Quantum computation and decision trees,''
{\em Physical Review A} {\bf 58} (1998), 915-928.

\bibitem{godsil08}
Chris Godsil,
``Periodic Graphs,''
math/08062074.

\bibitem{gr}
Chris Godsil and Gordon Royle,
{\em Algebraic Graph Theory},
Springer, 2001.

\bibitem{k06}
Viv Kendon,
``Quantum walks on general graphs,''
{\em International Journal of Quantum Information} {\bf 4}:5 (2006), 791-805.

\bibitem{niven}
Ivan Niven,
{\em Irrational Numbers},
The Mathematical Association of America, 1965.

\bibitem{s05}
Wasin So,
``Integral circulant graphs,''
{\em Discrete Mathematics} {\bf 306}:153-158, 2005.

\bibitem{sss07}
Nitin Saxena, Simone Severini, and Igor Shparlinski,
``Parameters of Integral Circulant Graphs and Periodic Quantum Dynamics,''
{\em International Journal of Quantum Information} {\bf 5}(3):417-430, 2007.

\bibitem{sw78}
Allen J. Schwenk and Robin J. Wilson,
``Eigenvalues of Graphs,''
in {\em Selected Topics in Graph Theory}, Lowell W. Beineke and Robin J. Wilson (eds.),
Academic Press (1978), 307-336.

\bibitem{tpvh08}
Dimitris I. Tsomokos, Martin B. Plenio, In\'{e}s de Vega, and Susana F. Huelga,
``State transfer in highly connected networks and a quantum Babinet principle,''
{\em Physical Review A} {\bf 78} (2008), 062310.

\end{thebibliography}
\end{document}